\documentclass[conference,letterpaper]{IEEEtran}
\usepackage[utf8]{inputenc}
\usepackage{amsthm, amssymb, amsfonts, amsmath}
\usepackage{graphicx}

\newtheorem{theorem}{Theorem}[section]

\newtheorem{corollary}[theorem]{Corollary}
\newtheorem{definition}[theorem]{Definition}

\newtheorem{lemma}[theorem]{Lemma}

\newtheorem{claim}[theorem]{Claim}

\DeclareMathOperator*{\Ber}{Ber}

\newcommand{\kk}{{\mathbf{k}}}

\newcommand{\R}{{\mathbb{R}}}

\newcommand{\Q}{{\mathbb{Q}}}

\newcommand{\Z}{{\mathbb{Z}}}
\newcommand{\ZZ}{\Sigma_2}

\newcommand{\U}{\mathcal{U}}

\newcommand{\X}{\mathcal{X}}

\newcommand{\SNR}{\text{\normalfont SNR}}

\begin{document}

\title{Sample Complexity of the Boolean Multireference Alignment Problem}

\author{\IEEEauthorblockN{Emmanuel Abbe}
\IEEEauthorblockA{Princeton University\\
\texttt{eabbe@princeton.edu}}
\and
\IEEEauthorblockN{João Pereira}
\IEEEauthorblockA{Princeton University\\
\texttt{jpereira@princeton.edu}}
\and
\IEEEauthorblockN{Amit Singer}
\IEEEauthorblockA{Princeton University\\
\texttt{amits@math.princeton.edu}}}

\maketitle

\begin{abstract}
The Boolean multireference alignment problem consists in recovering a Boolean 
signal from multiple shifted and noisy observations. In this paper we obtain an 
expression for the error exponent of the maximum A posteriori decoder. This 
expression is used to characterize the number of measurements needed for signal 
recovery in the low SNR regime, in terms of higher order autocorrelations of 
the signal. The characterization is explicit for various signal dimensions, 
such as prime and even dimensions. 
\end{abstract}

\section{Introduction}

The Boolean multireference alignment (BMA) problem consists of estimating an unknown signal $x\in \Z_2^L$, from noisy cyclically shifted copies $Y_1,\dots,Y_N\in \Z_2^L$, i.e.,
\begin{equation}\label{BMRADef}
Y_i=R^{S_i}x \oplus Z_i, \,i\in\{1,\dots,N\},
\end{equation}
where the error $Z_i\sim \Ber(p)^L$, the product measure of $L$ Bernoulli variables with parameter $p$, $\oplus$ denotes addition mod $2$, $R$ is the index cyclic shift operator that shifts a vector one element to the right $(x_1,\dots,x_N)\mapsto(x_N,x_1,\dots,x_{N-1})$, $R^{S_i}$ corresponds to applying $S_i$ times the operator $R$ and the shifts $S_i\sim\U(\Z_L)$, the uniform distribution in $\Z_L$.

The motivation to study this problem comes from the classical multireference alignment problem, where the signal and observations are real valued vectors, and the error is Gaussian white noise. Several algorithms were recently proposed to solve the problem, including angular synchronization \cite{angularsynchronization}, semidefinite program relaxations of the maximum likelihood decoder \cite{BCSZ} and reconstruction using the bispectrum \cite{sadler1992shift}. This problem is also an instance of a larger class of problems, called Non-Unique Games, which also includes the orientation estimation problem in cryo-electron microscopy \cite{NUG}.

Despite these advancements in algorithmic development, not much progress has 
been made in understanding the fundamental limits of signal recovery. The 
recent paper \cite{aguerrebere2016fundamental} investigated fundamental limits 
of shift recovery in multireference alignment, but not those of signal 
recovery. We note that estimating the shifts is impossible at low 
signal-to-noise ratio (SNR) even if an oracle presents us with the true signal. 
Also, the goal of many applications is signal recovery rather than shift 
estimation. Our paper aims to fill the gap on signal recovery, by studying the 
Boolean case. We show here that signal recovery is possible at arbitrarily low 
SNR, if sufficiently many measurements are available, and quantify this 
tradeoff. We do not consider here the problem of determining the sample 
complexity of multireference alignment in the real-valued Gaussian noise case, 
which is a topic of 
ongoing research \cite{OptRateMRA,SampCompMRA}.

In BMA the search space is finite, and the maximum A posteriori decoder (MAP) minimizes the probability of error. Our main contribution is an expression for the error exponent of MAP, in the low SNR regime, given in Theorems \ref{maintheorem} and \ref{corollary}. Our results imply how many measurements are needed, as a function of the SNR, in order to accurately estimate the signal.

The expression depends on the autocorrelations of the signal, defined in (\ref{autocorrelation}). Our results connect the order of autocorrelations needed to reconstruct the signal to the number of measurements needed to estimate the signal. This has some connections with previous theoretical work on uniqueness of the bispectrum \cite{Kakarala}.

We also consider some generalizations of the original problem in order to model some aspects of multireference alignment that arise in applications, such as the introduction of deletions.

\section{BMA Problem}

In the BMA problem, the errors are i.i.d. Bernoulli of parameter $p$. If $p=\frac12$, then the observations $Y_i\sim \Ber(\frac12)^L$, regardless of the original signal, and signal recovery is impossible. This corresponds to the case when $\SNR=0$. On the other hand, $p=0$ or $1$ corresponds to the noiseless case. Thus we define
\begin{equation}\label{SNR}
\SNR:=\left(p-\frac12\right)^2.
\end{equation}

In contrast to proposing an algorithm to solve the BMA problem, our paper focuses on its sample complexity, in the regime when $p\rightarrow\frac12$ and $\SNR\rightarrow 0$.

Note that the observations $Y_i$, $i\in [N]$, given the signal $x$, are i.i.d., since both the shifts $S_i$ and the errors $Z_i$ are i.i.d. For that reason we will drop the index $i$ when it is more convenient. We rewrite (\ref{BMRADef}), denoting by $x(j)$ the $j$-th entry of $x$.
\begin{equation}\label{BMRADefAlt}
Y(j)=x(S+j) \oplus Z(j),\, j\in \Z_L,
\end{equation}
where '$+$' is addition mod $L$.

Our paper also considers the sample complexity of the following variations of the basic BMA problem:

\begin{itemize}
	\item \emph{BMA Problem with consecutive deletions:}
	In this case the measurements $Y_1, \dots, Y_N$ are in $\Z_2^K$, with $K\le L$, and
	\begin{equation}\label{BMRAConDelDef}
	Y(j)=x(S+j) \oplus Z(j),\, j\in \Z_K.
	\end{equation}
	When $K=L$ we obtain the original BMA problem.	
	
	\item \emph{BMA Problem with known deletions:}
    Let $V\subset \Z_L$ be an ordered set of non-deletions, i.e. the set of deletions is $\Z_L\backslash V$. Now the measurements $Y_1, \dots, Y_N$ are in $\Z_2^K$, with $K=|V|$, and:
	\begin{equation}\label{BMRADelDef}
	Y(j)=x(S+V_j) \oplus Z(j),\, \forall j\in \Z_K,
	\end{equation}
	where $V_j$ denotes the $j$-th element of $V$. When $V=[K]$ we recover the BMA problem with consecutive deletions. 
	
	\item \emph{BMA Problem (and variations) with non uniform rotations:}
	Similar to the previous problems, but now the shifts follow some distribution $\xi$ in $\Z_L$.
\end{itemize}

These variations are motivated by problems similar to multireference alignment. The case of possible deletions is intended to model instances where the observations are only partial, whereas the extension to non-uniform shifts attempts to represent a non-symmetric version of the problem.

\section{Results}
We start by introducing the following notion of autocorrelation of a signal that is central to our main results.
\begin{definition}
The $(\xi,\kk)$-autocorrelation of $x$, with respect to a distribution $\xi$ in $\Z_L$ and $\kk=(k_1,k_2,\dots,k_d)\in \Z^d_L$ is defined as
\begin{equation}\label{autocorrelation}
A_{\xi,\kk}(x):=\sum _{s=1}^L \xi(s)x(k_1+s) \cdots x(k_d+s).
\end{equation}
We refer to $d=|k|$ as the order of the auto-correlation. When $\xi\sim \U(\Z_L)$, we simply write $\kk$-autocorrelation and $A_{\kk}$. Notice $A_\kk$ is shift invariant, that is $A_{\kk}(x)=A_{\kk}(R^s x)$, and in this case we may assume $k_1=0$.

We define the minimum autocorrelation order necessary to distinguish $x_1$ and $x_2$ under $\xi$ and $V$ as
\begin{equation}\label{mincorrorder}	t_{\xi,V}(x_1,x_2):=\inf\{d:A_{\xi,\kk}(x_1)\neq A_{\xi,\kk}(x_2),\kk\in V^d\},
\end{equation}
where $V^d$ denotes the vectors in $Z_2^d$ with entries in $V$. The minimum autocorrelation order necessary to describe all signals in $\X$ is defined as
\begin{equation}\label{mincorrorderX}
t_{\xi,V}(\X):=\max_{\substack{x_1,x_2\in \X\\x_1\neq x_2}}t_{\xi,V}(x_1,x_2).
\end{equation}
\end{definition}

Given a prior distribution on the signals $P_X$, with support $\X$, denote by $X$ the random variable with distribution $P_X$. Given an algorithm for BMA the probability of error is defined as
\begin{equation}\label{ProbError}
P(\hat X \neq X)=\sum_{x_i\in\X}P(\hat{X}\neq x_i)P_X(x_i),
\end{equation}
where $\hat X$ is the answer given by the algorithm. In the BMA problem the search space is finite, thus MAP minimizes the probability of error (\ref{ProbError}). We obtain results that do not depend on the prior distribution, they depend only on its support.
 
\begin{theorem}\label{maintheorem}
Consider the BMA problem with known deletions $Z_L\backslash V$ and shift distribution $\xi$. Let $\X\subset Z_2^L$ be the support of the prior distribution of the signals and $\mu_{x}$ the conditional distribution in $\Z_2^K$ of the observations $Y$ given the signal $x$, where $K=|V|$. The probability of error of the MAP estimator, denoted by $P_e$, has the following asymptotic behavior
\begin{equation} \label{SanovConst}
\lim_{N\rightarrow \infty} \frac{1}{N} \log P_e = \min_{\substack{x_1,x_2\in \X\\x_1 \neq x_2}}C(\mu_{x_1},\mu_{x_2}),
\end{equation}
with
\begin{multline}
C(\mu_{x_1},\mu_{x_2})=\\
\frac{2^{4t-3}}{t!}\SNR^{t} \sum _{\kk\in V^t}
\Big(A_{\xi,\kk}(x_1)-A_{\xi,\kk}(x_2)\Big)^2 + O(\SNR^{t+1}), \label{Chernoffexpansion}
\end{multline}
and $t=t_{\xi,V}(x_1,x_2)$.
\end{theorem}
The theorem implies that the exponent on $\SNR$ is $t_{\xi,V}(\X)$. 
In the original problem, with uniform shifts and no deletions, the recovery of the original signal is possible only up to a shift, i.e. we can only recover $R^k x$, where $x$ is the original signal, and $k$ is some shift in $\Z_L$. For that reason, we consider $\X$ to have exactly one element of each class of all the shifts of a signal, i.e., there are no two elements in $\X$ where one is a shift of the other (for example, if $L$ is prime, then there are $2^L-2$ such elements).
\begin{corollary}\label{corollary}
Consider the original problem, with $V=[L]$, $\xi\sim\U(\Z_L)$ and $\X$ as defined above.  By inspection one can obtain the error exponent for $L\le 5$. For $L\ge 6$, we either have
\begin{equation}\label{corollaryeq}
\lim_{N\rightarrow \infty} \frac{1}{N} \log P_e =
\left\{\begin{array}{l}
\medskip \frac{2^{10}}{L}\SNR^{3} + O(\SNR^{4})\\
O(\SNR^{4})
\end{array}\right.
\end{equation}
Also, the first case occurs when $L$ is prime, and the second when $L\ge12$ and is even. The other values of $L$ remain open.
\end{corollary}

\section{Proof Techniques}

\begin{IEEEproof}[Proof of Theorem \ref{maintheorem}]
The proof consists of two main parts. The next theorem gives a formula to the error exponent and claim \ref{EqClaims} makes the connection with autocorrelations.

\begin{theorem}
	Consider the BMA problem with known deletions $Z_L\backslash V$ and shift distribution $\xi$. Let $\X\subset Z_2^L$ be the space of possible signals and $\mu_{x}:=P_{Y|X}(\cdot|x)$ the conditional distribution in $\Z_2^K$ of the observations given the signal $x$. The probability of error of the MAP estimator  $(P_e)$ has the following asymptotic behavior
	\begin{equation}\label{SanovTheoremEq}
		\lim_{N\rightarrow \infty} \frac{1}{N} \log P_e = \min_{x_1\neq x_2\in \X}C(\mu_{x_1},\mu_{x_2}),
	\end{equation}
	with
	\begin{multline}\label{sanovtaylorn}
	C(\mu_{x_1},\mu_{x_2})=\\
	\frac{\left(\frac12-p\right)^{2s}}{8(s!)^2} \sum_{y\in \Z_2^K} \frac{\left(\mu_{x_1}^{(s)}\left(y;\frac12\right)-\mu_{x_2}^{(s)}\left(y;\frac12\right)\right)^2}{\mu_{x_1}\left(y;\frac12\right)}\\
	+O\left(\frac12-p\right)^{2s+2},
	\end{multline}
	where $\mu_{x}^{(m)}(y;p)$ denotes the $m$-th derivative of $\mu_{x}(y;p)$ in $p$, i.e. the derivative of the conditional distribution in $y$ given $x$ in order of the Bernoulli parameter $p$, and 	
	$$s(x_1,x_2):=\inf\left\{m:\mu_{x_1}^{(m)}\left(y;\frac12\right)\neq \mu_{x_2}^{(m)}\left(y;\frac12\right),y\in \Z_2^K\right\}.$$

\end{theorem}

This theorem follows from Theorems $1$ and $2$ in \cite{NoisySanov}. Theorem $1$ is a corollary of Sanov Theorem \cite{CoverThomas}, which leads to (\ref{SanovTheoremEq}). However the expression obtained by Theorem $1$ is rather complex and not very interpretable. In Theorem $2$ \cite{NoisySanov} we Taylor expand (\ref{SanovTheoremEq}) and obtain a useful characterization in instances where the SNR is small. We use this expression to obtain (\ref{sanovtaylorn}).

\begin{claim}\label{EqClaims}
	If $\mu_{x_1}^{(m)}\left(y;\frac12\right)= \mu_{x_2}^{(m)}\left(y;\frac12\right)$ for all $m<n$ and $y\in\Z_2^K$, then the following expressions are equal:
	\begin{equation}\label{claimexpr1}
	\sum_{y\in \Z_2^K} \frac{\left(\mu_{x_1}^{(n)}\left(y;\frac12\right)-\mu_{x_2}^{(n)}\left(y;\frac12\right)\right)^2}{\mu_{x_1}\left(y;\frac12\right)}
	\end{equation}
	and
	\begin{equation}\label{claimexpr3}
	2^{4n}n!\sum _{\kk \in V^L}
	\Big(A_{\xi,\kk}(x_1)-A_{\xi,\kk}(x_2)\Big)^2.
	\end{equation}
\end{claim}
In fact, since the expressions $(\ref{claimexpr1})$ and $(\ref{claimexpr3})$ are both sum of squares, the claim implies that $t_{\xi,V}(x_1,x_2)=s(x_1,x_2)$, what concludes the proof of theorem \ref{maintheorem}.
\end{IEEEproof}

\begin{IEEEproof}[Proof of Claim \ref{EqClaims}]
Denote by $x(V)$ the vector in $\Z_2^K$ ($K=|V|$) that consists of the values of $x$ with indices in $V$, i.e. the $j$-th element of $x(V)$ is $x(V_j)$. Also, given $s\in \Z_L$ denote by $s+V$ the ordered set corresponding to the sum of each element in $V$ with $s$ mod $L$. Equation (\ref{BMRADelDef}) can then be rewritten, as
\begin{equation}\label{setsumnotation}
Y=x(S+V)\oplus Z
\end{equation}
Then since $Z\sim \Ber(p)^L$, we have
\begin{equation*}
\mu_{x}(y;p|S=s)=(1-p)^{K-w(y\oplus x(s+V))}p^{w(y\oplus x(s+V))},
\end{equation*}
where $w$ denotes the Hamming weight, and since $S\sim \xi$
\begin{equation}\label{muxdef1}
\mu_{x}(y;p)=\sum \limits_{s=1}^L \xi(s) (1-p)^{K-w(y\oplus 
x(s+V))}p^{w(y\oplus x(s+V))}.
\end{equation}
In the statement of the theorem we have $x\in \Z_2^L$, however it is convenient for the proof to consider the entries of $x$ to be $-1,1$, changed by the rule: $a\mapsto 1-2a$. We will call 
\begin{equation}\label{udef}
u:=1-2x\in \ZZ^L
\end{equation}
 the corresponding element of $x$ with $\pm 1$ values, where $\ZZ:=\{-1,1\}$, and $v:=1-2y$. In analogy to the Hamming weight, we define
\begin{equation}\label{Wdef}
W(u):=\sum _{s=1}^L u(s)=L-2w(x).
\end{equation}
With this we rewrite (\ref{muxdef1})
\begin{equation}\label{muxdef2}
\mu_{u}(v;p)=\sum \limits_{s=1}^L \xi(s) (1-p)^{\frac{K}{2}+\frac{W(v\oplus u(s+V))}{2}}p^{\frac{K}{2}-\frac{W(v\oplus u(s+V))}{2}},
\end{equation}
where $\mu_{u}(v;p):=\mu_{x}(y;p)$. For simplicity of notation denote
$$W_{v,u,s}:=W(v\oplus u(s+V)).$$
The claim is now proved by induction on $n$. By properties of Jacobi polynomials \cite{szeg1939orthogonal} we have
$$\left(p^{\frac{K}2-\frac{b}2}(1-p)^{\frac{K}2+\frac{b}2}\right)^{(m)}_{\left|p=\frac12\right.}=(-2)^{m-K}P_m(b),$$
where $P_m$ is a polynomial with the following property
\begin{equation}\label{polyleadcoeff}
P_m(b)=b^m+Q_m(b),
\end{equation}
where $Q_m$ has degree at most $m-1$, and $Q_0\equiv Q_1\equiv 0$.
Thus
\begin{equation}\label{mupoly}
\mu_{u}^{(m)}\left(v;\frac12\right)=
(-2)^{m-K}\sum \limits_{s=1}^L \xi(s) P_m(W_{v,u,s}).
\end{equation}
Then when $m=1$
\begin{multline*}
\sum_{v\in \ZZ^K}
\frac{\left(\mu^{(1)}_{u_1}\left(v;\frac12\right)-\mu^{(1)}_{u_2}\left(v;\frac12\right)\right)^2}{\mu_{u_1}\left(v;\frac12\right)}\\
=2^{2-K} \sum_{v\in \ZZ^K}\left[\sum \limits_{s=1}^L \xi(s) \left(W_{v,u_1,s}-W_{v,u_2,s}\right)\right]^2.
\end{multline*}
Now, by the induction hypothesis if $\mu_{u_1}^{(k)}\left(v;\frac12\right)=\mu_{u_2}^{(k)}\left(v;\frac12\right)$ for all $k\le n-1$, $v \in \ZZ^K$
$$\sum_{s=1}^L\xi(s)Q_n(W_{v,u_1,s})=\sum_{s=1}^L\xi(s)Q_n(W_{v,u_2,s}),$$
for all $v \in \ZZ^K$ since $Q_n$ has degree at most $n-1$. Thus by (\ref{polyleadcoeff}) and (\ref{mupoly})
\begin{multline}\label{claimequation1}
\sum_{v\in \ZZ^K} \frac{\left(\mu_{u_1}^{(n)}\left(v;\frac12\right)-\mu_{u_2}^{(n)}\left(v;\frac12\right)\right)^2}{\mu_{u_1}\left(v;\frac12\right)}=\\
2^{2n-K}\sum_{v\in \ZZ^K}\left[\sum \limits_{s=1}^L \xi(s) \left(W_{v,u_1,s}^n-W_{v,u_2,s}^n\right)\right]^2
\end{multline}
Now splitting the square of the sum on the RHS into a product of two sums and expanding, we obtain terms of the form
\begin{equation}\label{sumy}
\sum_{s_1=1}^L \sum_{s_2=1}^L \xi(s_1)\xi(s_2) (-1)^{\alpha+\beta} \sum_{v\in \ZZ^K} W_{v,u_\alpha,s_1}^n W_{v,u_\beta,s_2}^n,
\end{equation}
where $\alpha$ and $\beta$ are $1$ or $2$. By Lemma \ref{Wcorr} we get
\begin{multline}
\sum_{v\in \ZZ^K} W_{v,u_\alpha,s_1}^n W_{v,u_\beta,s_2}^n =\\
 2^K \sum_{\substack{A\in M_{[2n]}\\A\text{ is even}}}
C_A\prod_{i=1}^{|A|}\left(\sum_{k=1}^K\prod_{j=1}^{|a_i|}u_{a_{ij}}(k)\right),
\end{multline}
Where $u_{a_{ij}}$ is $u_\alpha(s_1+V)$ if $a_{ij}\le n$, and is $u_\beta(s_2+V)$ otherwise. So, since $|a_i|$ is even, as $A$ is an even partition, and the entries of $u_{a_{ij}}$ are $\pm1$,
$$\sum_{k=1}^K\prod_{j=1}^{|a_i|}u_{a_{ij}}(k)=\sum_{k\in V} u_\alpha(s_1+k)u_\beta(s_2+k)$$
if $|a_i \cap [n]|$ is odd, and it is $K$ otherwise. Then
\begin{multline*}
\sum_{v\in \ZZ^K} W_{v,u_\alpha,s_1}^n W_{v,u_\beta,s_2}^n=
\\R_n\left(\sum_{k\in V} u_\alpha(s_1+k)u_\beta(s_2+k)\right),
\end{multline*}
where $R_n$ is a polynomial with degree $n$ (with coefficients possibly depending on $K$ and $n$), and $R_1(b)= 2^k b$. It cannot have degree $n+1$ since $|A|\le n$, since it is an even partition of $[2n]$. For it to be a power of order $n$, we need $|A|=n$, so $|a_i|=2$ for $i=1,\dots,n$, thus $C_A=1$, by the Lemma. Also $|a_i \cap [n]|$ must be odd for all $i$, thus $|a_i\cap [n]|=1$. There are exactly $n!$ partitions with this property, so the leading coefficient of $R_n$ is $2^K n!$.  We also have
\begin{align}
\nonumber &\sum_{s_1=1}^L \sum_{s_2=1}^L \xi(s_1)\xi(s_2) \left(\sum_{k\in V} u_\alpha(s_1+k)u_\beta(s_2+k)\right)^n\\
\nonumber &=\sum_{s_1=1}^L \sum_{s_2=1}^L \xi(s_1)\xi(s_2) \sum_{\kk\in V^n} \prod_{i=1}^n u_\alpha(s_1+k_i)u_\beta(s_2+k_i)\\
\label{acorreq}&=\sum_{\kk\in V^n} A_{\xi,\kk}(u_\alpha)A_{\xi,\kk}(u_\beta),
\end{align}
Mimicing the argument used in (\ref{claimequation1}), the equation will be true 
for $n=1$, since $R_1(b)= 2^k b$, and by the induction hypothesis only the 
leading coefficient of $R_n$ is of interest, since the other terms will cancel 
with each other. We get
\begin{multline}\label{claimequation2}
\sum_{v\in \ZZ^K}\left[\sum \limits_{s=1}^L \xi(s) \left(W_{v,u_1,s}^n-W_{v,u_2,s}^n\right)\right]^2=\\
2^k n! \sum_{\kk\in V^n} \left(A_{\xi,\kk}(u_1)-A_{\xi,\kk}(u_2)\right)^2
\end{multline}
Now through some algebraic manipulation, and using again the argument of the leading coefficient, if $|\kk|= n$, then
\begin{multline}\label{uxcorreq}
\sum_{\kk\in V^n} \left(A_{\xi,\kk}(u_1)-A_{\xi,\kk}(u_2)\right)^2=\\
2^{2n}\sum_{\kk\in V^n} \left(A_{\xi,\kk}(x_1)-A_{\xi,\kk}(x_2)\right)^2
\end{multline}
This together with (\ref{claimequation1}) and (\ref{claimequation2}) concludes the proof.
\end{IEEEproof}

\begin{lemma}\label{Wcorr}
For any partition $A=\{a_1,\dots,a_{|A|}\}$ of the set $\{1,2,\dots,m\}$, denote by $a_{ij}$ the $j$-th entry of $a_i$ and $M_{[m]}$ the set of all such partitions. If $u_1,\dots, u_m\in \ZZ^K$
\begin{multline}\label{WcorrEq}
\sum_{v\in\ZZ^K}W(u_1\oplus v)\cdots W(u_m\oplus v)=
\\2^K \sum_{\substack{A\in M_{[m]}\\A\text{ is even}}}
C_A\prod_{i=1}^{|A|}\left(\sum_{k=1}^K\prod_{j=1}^{|a_i|}u_{a_{ij}}(k)\right),
\end{multline}
where $A$ is even if all $|a_i|$ are even for $i\in \{1,\dots,|A|\}$. Moreover, $C_A$ is a constant that depends only on the partition $A$ and is always $1$ if $|a_i|=2$ for all $i\in \{1,\dots,|A|\}$.
\end{lemma}
\begin{IEEEproof}
Recall (\ref{Wdef}). We have $W(u\oplus v)=\sum \limits_{k=1}^K u(k)v(k)$
\begin{align}
\nonumber &\sum_{v\in\ZZ^K}W(u_1\oplus v)\cdots W(u_m\oplus v)\\
\nonumber &=\sum_{k_1=1}^K\cdots\sum_{k_m=1}^K u_1(k_1)\cdots 
u_m(k_m)\sum_{v\in\ZZ^K}v(k_1)\cdots v(k_m)\\
\label{alldistinct} &=\sum_{A\in M_{[m]}}
\sum_{\substack{k_1,\dots,k_{|A|}=1\\\text{all 
distinct}}}^K\prod_{i=1}^{|A|}
\prod_{j=1}^{|a_i|}u_{a_{ij}}(k_i)\sum_{v\in\ZZ^K}\prod_{i=1}^{|A|}v(k_i)^{|a_i|},
\end{align}
The last sum is $2^K$ when $A$ is even, and $0$ otherwise. Using a 
combinatorial argument we can rewrite (\ref{alldistinct}) without the 
'all-distinct' condition, at the cost of a constant $C_A$, which is $1$ when 
$|a_i|=2$ for $i\in\{1,\dots,|A|\}$. We get
\begin{multline*}
2^K \sum_{\substack{A\in M_{[m]}\\A\text{ is even}}}
C_A\sum_{k_1,\dots,k_{|A|}=1}^K\prod_{i=1}^{|A|}\prod_{j=1}^{|a_i|}u_{a_{ij}}(k_i)=\\
=2^K \sum_{\substack{A\in M_{[m]}\\A\text{ is even}}}
C_A\prod_{i=1}^{|A|}\left(\sum_{k=1}^K\prod_{j=1}^{|a_i|}u_{a_{ij}}(k)\right)
\end{multline*}
\end{IEEEproof}

\begin{IEEEproof}[Proof of Corollary \ref{corollary}]
We first prove equation (\ref{corollaryeq}). Recall (\ref{autocorrelation}), and denote by
$$B_m(x_1,x_2):=\sum _{\kk\in 
\Z_L^m}\Big(A_{\kk}(x_1)-A_{\kk}(x_2)\Big)^2$$

and
$$B_m(L):=\min_{x_1\neq x_2\in \X} B_m(x_1,x_2)$$
Note that $B_m(x_1,x_2)=0$ if $m<t_{\xi,V}(x_1,x_2)$ by (\ref{mincorrorder}). 
For convenience let  $B(x_1,x_2):=B_{t_{\xi,V}(x_1,x_2)}(x_1,x_2)$ and $B(L):=B_{t_{\xi,V}(\X)}(L)$ . Using this notation we rewrite (\ref{SanovConst}) and  (\ref{Chernoffexpansion})
$$\lim_{N\rightarrow \infty} \frac{1}{N} \log P_e=B(L)\frac{2^{4t_L-3}}{t_L!}\SNR^{t_L}+O\left(\SNR^{t_L+1}\right)$$

Now equation (\ref{corollaryeq}) is equivalent to having $t_{\xi,V}(\X)\ge 3$  and $B_3(L)$ either $\frac{12}{L}$ or $0$. Turns out, for $L\ge 6$, if we take $$x^*_1=(1,1,0,1, \underbrace{0,\dots,0}_{L-4\text{ zeros}})\text{ and } x^*_2=(1,0,1,1,\underbrace{0,\dots,0}_{L-4\text{ zeros}}),$$ then $t_{\xi,V}(\X)\ge t_{\xi,V}(x^*_1,x^*_2)= 3$ and $B_3(L)\le B(x^*_1,x^*_2)=\frac{12}{L}$. Also we cannot have $\frac{12}{L}>B_3(L)>0$. This implies there exists $x_1$ and $x_2$ in $\X$ such that $\frac{12}{L}>B(x_1,x_2)>0$. Since it is positive, there is $\kk^* \in \Z_L^3$ such that $A_{\kk^*}(x_1)\neq A_{\kk^*}(x_2)$. But by definition (\ref{autocorrelation}), since $\xi(s)=\frac1L$, $L A_{\kk^*}(x)$  is an integer for $x\in \Z_2^L$, and $L^2 (A_{\kk^*}(x_1)-A_{\kk^*}(x_2))^2 \in \Z$.

Now by the definition we also have $A_{\sigma(\kk^*)}(x)=A_{\kk^*}(x)$, where $\sigma$ permutes the entries of $\kk^*$. Also, for $s\in \Z_L$, let $s+\kk^*:=(s+k^*_1,s+k^*_2,s+k^*_3)$, then $A_{s+\kk^*}(x)=A_{\kk^*}(x)$.
There is $6$ permutations and $L$ possible values for $s\in \Z_L$, so $B(x_1,x_2)$ is an integer multiple of $\frac{6}{L}$. (we can also have not trivial $s$ and $\sigma$ such that $s+\kk^*=\sigma(\kk^*)$ but that case also has the property mentioned). However we cannot have $B(x_1,x_2)=\frac{6}{L}$. That means there exists only one $\kk^*\in \Z_L^3$ (with permutations and shifts) such that $A_{\kk^*}(x_1)\neq A_{\kk^*}(x_2)$. Then
\begin{equation}\label{6Lcontradiction}
\sum _{\kk\in \Z_L^3}A_{\kk}(x_1)-A_{\kk}(x_2)=6L (A_{\kk^*}(x_1)-A_{\kk^*}(x_2))\neq 0
\end{equation}
On the other hand
\begin{align*}
\sum _{\kk\in \Z_L^3}A_{\kk}(x_1)
&=\frac1L \sum _{s=1}^L\sum _{\kk\in \Z_L^3} x(k_1+s) x(k_2+s) x(k_3+s)\\
&=L^3A_0(x_1)^3,
\end{align*}
where $A_0$ denotes $\kk$-autocorrelation with $\kk=0$. Since $t_L>1$, $A_0(x_1)=A_0(x_2)$, so equation (\ref{6Lcontradiction}) must be $0$, and equation 
(\ref{corollaryeq}) follows by contradiction.
Now if $L\ge 12$ is even, choose
$$x^*_1=(1,1,0,\underbrace{1,\dots,1}_{\frac{L}{2}-3\text{ ones}},0,0,1,\underbrace{0,\dots,0}_{\frac{L}{2}-3\text{ zeros}})$$
and $x^*_2$ the vector obtained by reversing the entries of $x^*_1$. Since one is the reverse of the other, they have same $1$ and $2$ order autocorrelations. Recall (\ref{udef}) and (\ref{autocorrelation}) and notice that in this case both $A_{\kk}(u_1)$ and $A_{\kk}(u_2)$ are $0$ when $|\kk|$ is odd, since half of the signal is the symmetric of the other half, i.e. $u_1(\{1,\dots,\frac{L}2\})=-u_1(\{\frac{L}2+1,\dots,L\})$. Now because of (\ref{uxcorreq}) we have $A_{\kk}(x_1)=A_{\kk}(x_2)$ when $|\kk|=3$, so $t_L\ge 4$, and $B_3(L)=0$.

Finally, let $L\ge 6$ be prime. We prove by contradiction that $t_L=3$ and $B_3(L)=\frac{12}{L}$. If this is not true, then it exists $x^*_1$ and $x^*_2$ such that $t_{x^*_1,x^*_2}>3$, so
\begin{equation}\label{kakaralathmcond}
A_{\kk}(x^*_1)=A_{\kk}(x^*_2),\quad \kk\in \Z^n_L,n\le 3 
\end{equation}
By Theorem 2 of paper \cite{Kakarala}, if the Fourier coefficients of $x^*_1$ and $x^*_2$ are non-zero, then equation (\ref{kakaralathmcond}) implies one is a shift of the other. Denote by $\{r_j^1\}_{j\in\Z_L}$ and $\{r_j^2\}_{j\in\Z_L}$ the Fourier coefficients of $x^*_1$ and $x^*_2$, respectively, which are given by
\begin{align}
r_j^\alpha&=\frac1{\sqrt{L}}\sum_{s=1}^L x_\alpha(s)\omega_L^{-js},\quad \alpha\in \{1,2\},j \in \Z_L,\\
&=\frac1{\sqrt{L}}\sum_{s:x_\alpha(s)=1} \omega_L^{-js},
\end{align}
where $\omega_L$ is the $L$'th root of unity. $r_0^\alpha=0$ implies $x^*_\alpha$ only has zeros, and $r_j^\alpha$ is $0$ only if $w^{-j}_L$ is a root of the polynomial
\begin{equation}\label{coeffpoly}
\sum_{s:x_\alpha(s)=1} b^s
\end{equation}
However, since $L$ is prime, the minimal polynomial of $w^{-j}_L$ in $\Q[x]$, 
for $L>j>0$, is $1+x+\dots +x^{L-1}$ \cite{jacobson2012lectures}, so this 
polynomial must divide (\ref{coeffpoly}). Thus $x^*_1$ and $x^*_2$ must be the 
all zeros and all ones signals, but these signals also do not satisfy 
(\ref{kakaralathmcond}).
\end{IEEEproof}

\section*{Acknowledgment}
\addcontentsline{toc}{section}{Acknowledgment}

A. S. and J. P. were partially supported by Award Number R01GM090200 from the NIGMS, FA9550-12-1-0317 from AFOSR, Simons Foundation Investigator Award and Simons Collaborations on Algorithms and Geometry, and the Moore Foundation Data-Driven Discovery Investigator Award.

E. A. is partially supported by NSF CAREER Award CCF-1552131 and ARO grant W911NF-16-1-0051.



\end{document}